\documentclass[floatfix, prl, twocolumn, showpacs, showkeys, preprintnumbers, superscriptaddress]{revtex4-1}
\usepackage[utf8]{inputenc}
\usepackage{mathrsfs,bm}
\usepackage{natbib}
\usepackage{txfonts}
\usepackage{amssymb}
\usepackage{rotating}
\usepackage{indentfirst}
\usepackage{graphicx,booktabs}
\usepackage{multirow}
\usepackage{slashed}
\usepackage{overpic}
\usepackage{color}
\usepackage{amssymb}
\usepackage{hyperref}
\usepackage{bbding}
\usepackage{url}

\begin{document}
\title{Weak radiative hyperon decays in covariant baryon chiral perturbation theory}

\author{Rui-Xiang Shi}
\affiliation{School of Space and Environment,  Beihang University, Beijing 102206, China}
\affiliation{School of Physics, Beihang University, Beijing 102206, China}

\author{Shuang-Yi Li}
\affiliation{School of Physics,  Beihang University, Beijing 102206, China}

\author{Jun-Xu Lu}
\affiliation{School of Space and Environment, Beihang University, Beijing 102206, China}
\affiliation{School of Physics, Beihang University, Beijing 102206, China}

\author{Li-Sheng Geng}
\email[Corresponding author: ]{lisheng.geng@buaa.edu.cn}
\affiliation{School of
Physics,  Beihang University, Beijing 102206, China}
\affiliation{Beijing Key Laboratory of Advanced Nuclear Materials and Physics, Beihang University, Beijing 100191, China}
\affiliation{School of Physics and Microelectronics, Zhengzhou University, Zhengzhou, Henan 450001, China }

\begin{abstract}
Weak radiative hyperon decays,  important to test the strong interaction  and relevant in searches for beyond the standard model physics, have remained puzzling both experimentally and theoretically for a long time. The recently updated branching fraction  and first measurement of the asymmetry parameter of $\Lambda\to n\gamma$ by the BESIII Collaboration further exacerbate the issue, as none of the existing predictions can describe the data. We show in this letter that the covariant baryon chiral perturbation theory, with constraints from the latest measurements of hyperon non-leptonic decays, can well describe the BESIII data. The predicted branching fraction and asymmetry parameter for $\Xi^-\to\Sigma^-\gamma$ are also in agreement with the experimental data. We note that a more precise measurement of the asymmetry parameter, which is related with that of $\Sigma^+\to p\gamma$, is crucial to test Hara's theorem. We further predict the branching fraction and asymmetry parameter of $\Sigma^0\to n\gamma$, whose future measurement can serve as a highly nontrivial check on our understanding of weak radiative hyperon decay and on the covariant baryon chiral perturbation theory.

\end{abstract}


\maketitle
 {\it Introduction:} Weak radiative hyperon decays~(WRHDs) are interesting physical processes where the electromagnetic, weak, and strong interactions all play a role~\cite{Lach:1995we}. Despite of their deceptively simple two-body kinematics, studies of such decays have remained challenging both experimentally and theoretically. Recently, the BSEIII Collaboration reported a new measurement of the  $\Lambda\to n\gamma$ decay. The measured branching fraction and asymmetry parameter are~\cite{BESIII:2022rgl}:
\begin{eqnarray}
&&{\cal B}(\Lambda\to n\gamma)=(0.832\pm0.038\pm0.054)\times10^{-3},\\
&&\alpha_\gamma(\Lambda\to n\gamma)=-0.16\pm0.10\pm0.05,
\end{eqnarray}
where the first uncertainty is statistical and the second is systematic. The  branching fraction is only about one half of the current PDG average  $(1.75\pm0.15)\times10^{-3}$~\cite{ParticleDataGroup:2020ssz}. Compared with previous measurements~\cite{Biagi:1986vn,Noble:1992ya,Larson:1993ig}, the new measurement is featured by a larger statistics and a small uncertainty. Moreover, the asymmetry parameter $\alpha_\gamma$ is determined for the first time. As a result, it provides a highly nontrivial check on our understanding of the WRHDs and on the validity of various theoretical models. In particular, as stressed in Ref.~\cite{Zenczykowski:2020hmg}, the measurement of $\alpha_\gamma(\Lambda\to n\gamma)$ could yield a definite answer on the issue of Hara's theory~\cite{Hara:1964zz}.

The puzzle of  WRHDs started with the experimental measurement of a surprisingly  large asymmetry for $\Sigma^+\rightarrow p\gamma$~\cite{Gershwin:1969fpe}, which contradicts Hara’s theorem~\cite{Hara:1964zz}.  Up to now, five of the six WRHDs~\cite{Gershwin:1969fpe,Manz:1980td,Bristol-Geneva-Heidelberg-Lausanne-QueenMaryColl-Rutherford:1985ksh,
Kobayashi:1987yv,Hessey:1989ep,E761:1992atm,E761:1994yxs,Yeh:1974wv,James:1990as,NA48:1999dxg,NA48:2004gyg,
Batley:2010bp,Bensinger:1988va,Teige:1989uk,KTeV:2000dsr,Bristol-Geneva-Heidelberg-Lausanne-QueenMaryColl-Rutherford:1986kqv,
E761:1993unn,BESIII:2022rgl} have been measured except for that of $\Sigma^0\rightarrow n\gamma$. Various theoretical models have been developed to understand the experimental data (see Ref.~\cite{Lach:1995we} for earlier works before 1995). As the asymmetry parameters have changed dramatically~\footnote{It is interesting to note that the signs of the asymmetry parameters for  $\Xi^0\to\Lambda\gamma$ and $\Xi^0\to\Sigma^0\gamma$ were wrongly assigned in the first measurements~\cite{Teige:1989uk,James:1990as}, i.e., $\alpha_\gamma(\Xi^0\to\Sigma^0\gamma)=0.20\pm0.32\pm0.05$, $\alpha_\gamma(\Xi^0\to\Lambda^0\gamma)=0.43\pm0.44$. In 2000, the KTeV Collaboration reported a new measurement of the asymmetry parameter $\alpha_\gamma(\Xi^0\to\Sigma^0\gamma)=-0.63\pm0.09$~\cite{KTeV:2000dsr}. Three years later, the NA48 experiment at the CERN SPS also re-measured the asymmetry parameter $\alpha_\gamma(\Xi^0\to\Lambda\gamma)=-0.78\pm0.18\pm0.06$~\cite{NA48:2004gyg}. These two asymmetry parameters were revised in 2010 by the NA48/1 Collaboration~\cite{Batley:2010bp}, yielding $\alpha(\Xi^0\to\Lambda\gamma)=-0.704\pm0.019\pm0.064$  and $\alpha(\Xi^0\to\Sigma^0\gamma)=-0.729\pm0.030\pm0.076$. Compared with the previous measurements~\cite{Teige:1989uk,James:1990as}, the asymmetry parameters have changed significantly.}, particularly those of $\Xi^0\rightarrow\Lambda\gamma$ and $\Xi^0\rightarrow\Sigma^0\gamma$, some subsequent works~\cite{Dubovik:2008zz,Zenczykowski:2005cs,Niu:2020aoz,Wang:2020wxn,Zenczykowski:2020hmg} updated the early studies. We note that the phenomenological  model of Ref.~\cite{Dubovik:2008zz} and the broken SU(3) model of Ref.~\cite{Zenczykowski:2005cs}
are able to explain the experimental data at least qualitatively. However, their predictions for the $\Lambda\to n\gamma$ decay are quite different from the new BESIII data~\cite{BESIII:2022rgl}.

Chiral perturbation theory ($\chi{\rm PT}$) as the low energy effective theory of quantum chromodynamics (QCD) has been very successful in
studying low-energy strong interaction physics~\cite{Weinberg:1978kz,Gasser:1983yg,Gasser:1987rb,Weinberg:1990rz,Burdman:1992gh}. Utilizing chiral symmetry and its explicit and spontaneous breaking, $\chi{\rm PT}$ allows for a model independent description of the strong interaction which can be improved order by order and whose uncertainties can be quantified (see Ref.~\cite{Scherer:2012xha} for a pedagogical introduction). In the one-baryon sector, because of the large non-zero baryon mass in the chiral limit, additional care is needed to ensure a systematic power counting. Over the years, three approaches have been developed to tackle this issue, namely the so-called heavy baryon~(HB) $\chi{\rm PT}$~\cite{Jenkins:1990jv,Bernard:1995dp}, the infrared $\chi{\rm PT}$~\cite{Becher:1999he}, and the extended-on-mass-shell~(EOMS) $\chi{\rm PT}$~\cite{Gegelia:1999gf,Fuchs:2003qc}. For a short introduction to these three methods, see, e.g., Ref.~\cite{Geng:2013xn}. $\chi{\rm PT}$ has been applied to study the WRHDs  in the covariant formulation~\cite{Neufeld:1992np} and the heavy baryon formulation~\cite{Jenkins:1992ab,Bos:1994ed,Bos:1996ig,Bos:1996ja} up to one-loop order and at tree level but considering the contribution of heavier resonances~\cite{Borasoy:1999nt}. One should note that in Ref.~\cite{Neufeld:1992np}, the two relevant low energy constants~(LECs) $h_D$, $h_F$ and hyperon non-leptonic decay amplitudes  were determined by fitting to the 1992 PDG data~\cite{Jenkins:1991bt} and therefore need to be updated. In addition, no effort was taken to ensure a consistent power counting. In the present work, inspired by the latest experimental progress~\cite{BESIII:2022rgl,ParticleDataGroup:2020ssz,BESIII:2018cnd,Ablikim:2022ick,BESIII:2021ypr,Ireland:2019uja} and the theoretical developments in the formulation of a covariant baryon chiral perturbation theory~\cite{Becher:1999he,Gegelia:1999gf,Fuchs:2003qc} and its successful applications in the one-baryon sector~\cite{Geng:2008mf} and two-nucleon sector~\cite{Lu:2021gsb}, we revisit the WRHDs in covariant baryon chiral perturbation theory~(B$\chi{\rm PT}$) with the EOMS renormalization scheme.

{\it Theoretical framework:} The effective Lagrangian for the weak radiative hyperon decay $B_i\to B_f\gamma$ reads
\begin{eqnarray}
{\cal L}=\frac{eG_F}{2}\bar{B}_f(a+b\gamma_5)\sigma^{\mu\nu}B_iF_{\mu\nu},\label{WRHDs:lag}
\end{eqnarray}
 and the corresponding decay rate is (\ref{WRHDs:lag})
\begin{eqnarray}
&&\frac{d\Gamma}{d\cos\theta}=\frac{e^2G_F^2}{\pi}(|a|^2+|b|^2)[1+\frac{2{\rm Re}(ab^*)}{|a|^2+|b|^2}\cos\theta]\cdot|\vec{k}|^3,\nonumber\\
&&\alpha_\gamma=\frac{2{\rm Re}(ab^*)}{|a|^2+|b|^2},\qquad\Gamma=\frac{e^2G_F^2}{\pi}(|a|^2+|b|^2)\cdot|\vec{k}|^3,\label{WRHDs:asypara}
\end{eqnarray}
where $\alpha_\gamma$ is the asymmetry parameter, $G_F$ is the Fermi constant, and $\theta$ is the angle between the spin of the initial hyperon $B_i$ and the 3-momentum $|\vec{k}|=\frac{m_i^2-m_f^2}{2m_i}$ of the final baryon $B_f$. The amplitudes $a$ and $b$ are partity-conserving and partiy-violating pieces, respectively. In total, there are six decay channels for the WRHDs of the ground-state octet baryons, i.e., $\Lambda\to n\gamma$, $\Sigma^+\to p\gamma$, $\Sigma^0\to n\gamma$, $\Xi^0\to\Lambda\gamma$, $\Xi^0\to\Sigma^0\gamma$ and $\Xi^-\to\Sigma^-\gamma$.

\begin{figure*}[htpb]
  \centering
   \includegraphics[width=0.24\linewidth]{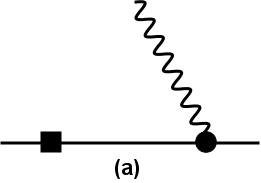}~~~
  \includegraphics[width=0.24\linewidth]{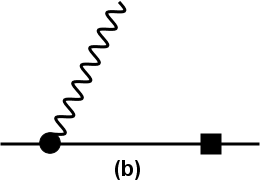}~~~
   \includegraphics[width=0.24\linewidth]{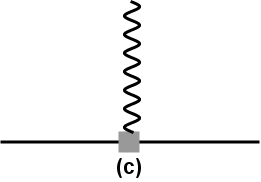}\\
  \includegraphics[width=0.24\linewidth]{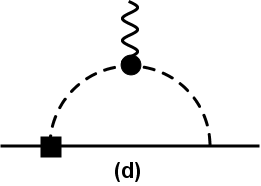}~~~
  \includegraphics[width=0.24\linewidth]{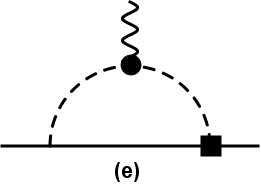}~~~
  \includegraphics[width=0.24\linewidth]{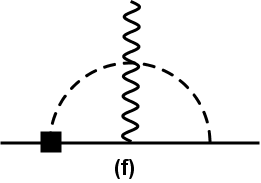}~~~
  \includegraphics[width=0.24\linewidth]{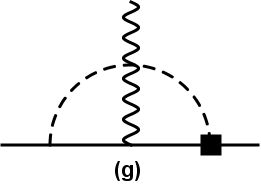}
  \caption{Feynman diagrams contributing to amplitudes $a$ and $b$ of the WRHDs. They can be classified into three groups. The pole diagrams (a) and (b) contribute at LO to the real part of amplitude $a$, and the direct photon emission diagram (c) contributes at NLO to the real parts of $a$ and $b$. The other diagrams contribute to amplitudes $a$ and $b$ at ${\cal O}(p^2)$. The solid, dashed, and wiggly lines represent octet baryons, Goldstone bosons, and photons, respectively. The heavy dots denote the ${\cal O}(p^2)$ vertices. The solid squares denote the LO weak interaction vertices, and the gray one arises at ${\cal O}(p^2)$. In diagrams (d,e,f,g), the  $P$- and $S$-wave weak vertices are of the same order, and contribute to the WRHDs at ${\cal O}(p^0)$.}\label{Fig1:FeynDiagrams}
\end{figure*}

In the present work, we calculate the branching fractions $\cal B$ and
asymmetry parameters $\alpha_\gamma$ of the WRHDs order by order.  In the B$\chi{\rm PT}$, the chiral order is defined as $n_\chi=4L-2N_M-N_B+\sum_kkV_k$ for a Feynman diagram with $L$ loops, $N_M~(N_B)$ internal meson~(baryon) propagators, and $V_k$ vertices from $k^{\rm th}$ order Lagrangians~\cite{Geng:2013xn}.

 Fig.~\ref{Fig1:FeynDiagrams} shows the Feynman diagrams contributing to the decay amplitudes $a$ and $b$ up to ${\cal O}(p^2)$, where we have ignored  the tiny short-distance contributions~\cite{Jenkins:1992ab,Neufeld:1992np}. The leading order (LO) contribution is determined by the following Lagrangians:
\begin{eqnarray}
&&{\cal L}_{\Delta S=1}^{(0)}=\sqrt{2}G_Fm_{\pi}^2F_{\phi}\langle h_D\bar{B}\{u^\dag\lambda u,B\}+h_F\bar{B}[u^\dag\lambda u,B]\rangle,\qquad\label{WRHDs:lagTree1}\\
&&{\cal L}_{MB}^{(2)}=\frac{b_6^D}{8m_B}\langle\bar{B}\sigma^{\mu\nu}\{F_{\mu\nu}^+,B\}\rangle+\frac{b_6^F}{8m_B}\langle\bar{B}\sigma^{\mu\nu}
[F_{\mu\nu}^+,B]\rangle,\label{WRHDs:lagTree2}
\end{eqnarray}
where $h_D$, $h_F$, $b_6^D$, and $b_6^F$ are LECs, the superscripts of the Lagrangian terms stand for the chiral order, $\sigma^{\mu\nu}=\frac{i}{2}[\gamma^\mu,\gamma^\nu]$, $F_{\mu\nu}^+=(u^\dag QF_{\mu\nu}u+uQF_{\mu\nu}u^\dag)$, $F_{\mu\nu}=\partial_\mu A_\nu-\partial_\nu A_\mu$,
$\lambda=(\lambda_6+i\lambda_7)/2$ and $Q=|e|{\rm diag}(2,-1,-1)/3$ are the $s\to d$ transition matrix and quark charge matrix, respectively, where $\lambda_6$ and $\lambda_7$ are the Gell-Mann matrices,  $u={\rm exp}[i\Phi/2F_\phi]$, with the unimodular matrix containing the pseudoscalar octet $\Phi$, and $F_\phi$ the pseudoscalar decay constant. In the present work, we take their physical values $F_\pi=92.4~{\rm MeV}$, $F_K=1.22F_\pi$, and $F_\eta=1.3F_\pi$. The LECs $h_D$ and $h_F$ are determined to be $-0.61(24)$ and $1.42(14)$,  by fitting to the latest experimental  hyperon non-leptonic decay data~\cite{ParticleDataGroup:2020ssz,Ablikim:2022ick,BESIII:2021ypr,Ireland:2019uja}, which is detailed in the Supplemental Material. The other LECs $b_6^D$ and $b_6^F$ are related to the magnetic moment of octet baryons. As a result, we can  use the experiment baryon magnetic moments instead of by fitting them to data~\cite{Geng:2008mf,Geng:2009hh}. For the masses of the octet baryons and mesons, we take the latest PDG  values~\cite{ParticleDataGroup:2020ssz}.

At $\mathcal{O}(p^2)$, the CPS symmetry~\footnote{CPS is CP followed by the SU(3) transformation of $u \rightarrow-u, d \rightarrow s, s \rightarrow d$ which exchanges $\mathrm{s}$ and $\mathrm{d}$ quarks~\cite{Bernard:1985wf}.} dictates the existence of five unknown LECs~\cite{Jenkins:1992ab},
\begin{eqnarray}
&&\mathcal{L}_\alpha^{(2)}=C_\alpha\langle\bar{B}\sigma^{\mu\nu}F_{\mu\nu} \lambda QB\rangle,\label{Eq:OP2alpha}\\
&&\mathcal{L}_\beta^{(2)}=C_\beta\langle\sigma^{\mu\nu}F_{\mu\nu} \bar{B}QB\lambda\rangle,\\
&&\mathcal{L}_\gamma^{(2)}=C_\gamma\langle\bar{B}\sigma^{\mu\nu}F_{\mu\nu}  B\lambda Q\rangle,\\
&&\mathcal{L}_\sigma^{(2)}=C_\sigma\langle\bar{B}\sigma^{\mu\nu}F_{\mu\nu}\lambda BQ\rangle,\label{Eq:OP2Sigma}\\
&&\mathcal{L}_\rho^{(2)}=C_\rho\left(\langle\bar{B}\sigma^{\mu\nu}\gamma_5F_{\mu\nu} Q\rangle\langle B\lambda\rangle-\langle\bar{B}\sigma^{\mu\nu}\gamma_5F_{\mu\nu}\lambda\rangle\langle BQ\rangle\right).\qquad\label{Eq:OP2rho}
\end{eqnarray}
It is easy to see that $\mathcal{L}_{\alpha,\beta,\gamma,\sigma}^{(2)}$ contribute to amplitude $a$, while  $\mathcal{L}_\rho^{(2)}$ contributes to amplitude $b$~\footnote{A few remarks are in order. First, a sixth counterterm appears in Ref.~\cite{Neufeld:1992np}, which can be easily shown to be redundant using the Cayley-Hamilton identity and noting that $\bar{B}$, $B$,
$Q$, and $\lambda$ are all traceless,  and $[\lambda,Q]=0$. Second, in Ref.~\cite{Bos:1996ig}, the authors introduced explicitly two terms at  $\mathcal{O}(p)$ that break the U-spin symmetry and contribute to the WRHDs of $\Xi^-$ and $\Sigma^+$. We donot introduce such terms in the present work.}.

At the same chiral order, the WRHDs also receive loop contributions  characterized by the lowest order Lagrangian ${\cal L}_{\Delta S=1}^{(0)}+{\cal L}_B^{(1)}+{\cal L}_{M}^{(2)}+{\cal L}_{MB}^{(1)}$, the latter three of which read,
\begin{eqnarray}
&&{\cal L}_B^{(1)}=\langle\bar{B}i\gamma^\mu D_\mu B-m_0\bar{B}B\rangle,\nonumber\\
&&{\cal L}_{M}^{(2)}=\frac{F_\phi^2}{4}\langle u_\mu u^\mu+\chi^+\rangle,\nonumber\\
&&{\cal L}_{MB}^{(1)}=\frac{D}{2}\langle\bar{B}\gamma^\mu\gamma_5\{u_\mu,B\}\rangle
+\frac{F}{2}\langle\bar{B}\gamma^\mu\gamma_5[u_\mu,B]\rangle,\label{WRHDs:lagloop}
\end{eqnarray}
with
\begin{eqnarray}
&&D_\mu B=\partial_\mu B+[\Gamma_\mu,B],~\chi^\pm=u^\dag\chi u^\dag\pm u\chi u\nonumber\\
&&\Gamma_\mu=\frac{1}{2}(u^\dag\partial_\mu u+u\partial_\mu u^\dag)-\frac{i}{2}(u^\dag v_\mu u+uv_\mu u^\dag)=-ieQA_\mu,\nonumber\\
&&u_\mu=i(u^\dag\partial_\mu u-u\partial_\mu u^\dag)+(u^\dag v_\mu u-uv_\nu u^\dag),
\end{eqnarray}
where $m_0=880$ MeV stands for the baryon mass in the chiral limit~\cite{Ren:2012aj}, $v_\mu$ is the vector source, $\chi=2B_0{\cal M}$ with ${\cal M}$ the quark mass matrix ${\cal M}={\rm diag}(m_q,m_q,m_s)$, and $B_0=|\langle0|\bar{q}q|0\rangle|/F_\phi^2$. In this work we take the axial vector couplings $D=0.793(18)$ and $F=0.476(17)$ determined from the semi-leptonic hyperon decays~(SHD)~\footnote{In our work, the central value of each axial vector coupling is the average of the two central values determined by fitting to the SHD data and SHD $+$ lattice QCD data in Ref.~\cite{Ledwig:2014rfa}. We treat the differences between the central values obtained by the two fittings as systematic uncertainties. The total uncertainties  are  the statistical and the systematic errors added in quadrature.}.

The pole diagrams (a) and (b)  contribute to the real part of amplitude $a$ at ${\cal O}(p)$, which can be easily obtained from Eqs.~(\ref{WRHDs:lagTree1}) and (\ref{WRHDs:lagTree2}). The five counter-terms of Eqs.~(\ref{Eq:OP2alpha}-\ref{Eq:OP2rho})   contribute to the real part of amplitudes $a$ and $b$ at ${\cal O}(p^2)$ via the direct photon emission diagram Fig.~(1c). The explicit expressions for these contributions  are given in the Supplemental Material.

The contributions to amplitude $b$ from loop diagrams Figs.~(1d), (1e), (1f) and (1g) are of ${\cal O}(p^2)$ and read,
\begin{eqnarray}
b_{B_iB_f}^{(\rm 2,loop)}&=&\sum_{B_j\phi}\frac{m_{\pi}^2}{F_\phi}\left(\xi_{B_j\phi}^{(d)}H_S^{(d)}(m_i,m_j,m_f,m_{\phi})\right.\nonumber\\
&+&\xi_{B_j\phi}^{(e)}H_S^{(e)}(m_i,m_j,m_f,m_{\phi})+\xi_{B_j\phi}^{(f)}H_S^{(f)}(m_i,m_j,m_f,m_{\phi})\nonumber\\
&+&\left.\xi_{B_j\phi}^{(g)}H_S^{(g)}(m_i,m_j,m_f,m_{\phi})\right),\label{Eq:loopampb}
\end{eqnarray}
where the coefficients $\xi_{B_j\phi}^{(d,e,f,g)}$ depend on the LECs $h_D$, $h_F$, $D$, and $F$. The explicit expressions of these coefficients and loop  functions $H_S^{(d,e,f,g)}(m_i,m_j,m_f,m_{\phi})$ can be found in the Supplemental Material.

It should be noted that one can only reliably determine the imaginary part of amplitude $a$ because of the long-standing $S/P$ puzzle of the hyperon non-leptonic decays~\cite{Holstein:2000yf} and the existence of four unknown LECs $C_{\alpha,\beta,\gamma,\sigma}$. Following Ref.~\cite{Jenkins:1992ab}, we  treat the real part of amplitude $a$ as a free parameter. The imaginary parts ${\rm Im}~a_{B_iB_f}^{(\rm 2,loop)}$ of amplitudes $a$ from loop contributions are given explicitly in the Supplemental Material.

{\it Results and discussions:} With constraints from the latest measurements of hyperon
non-leptonic decays~\cite{ParticleDataGroup:2020ssz,Ablikim:2022ick,BESIII:2021ypr,Ireland:2019uja}, we can obtain  the branching fractions and asymmetry parameters of the  WRHDs. 
We first compute the parity-conserving  and parity-violating  amplitudes $a$ and $b$. Up to ${\cal O}(p^2)$, the total amplitudes are a sum of the tree and loop contributions and read:
\begin{eqnarray}
&&a_{B_iB_f}=a_{B_iB_f}^{(\rm 1,tree)}+a_{B_iB_f}^{(\rm 2,tree)}+a_{B_iB_f}^{(\rm 2,loop)}={\rm Re}~a_{B_iB_f}+{\rm Im}~a_{B_iB_f}^{(\rm 2,loop)},\nonumber\\
&&b_{B_iB_f}=b_{B_iB_f}^{(\rm 2,tree)}+b_{B_iB_f}^{(\rm 2,loop)}.\label{Eq:TotalAmpab}
\end{eqnarray}
As mentioned above, we take  the real part of amplitude $a$ for each decay mode as a free parameter.

It is important to note that the only unknown LEC $C_\rho$  can be determined by fitting to the branching fractions ${\cal B}$ and asymmetry parameters $\alpha_\gamma$  of the $\Xi^0\to\Sigma^0 \gamma$ and $\Xi^0\to\Lambda\gamma$ decays, together with  ${\rm Re}~a_{\Xi^0\Lambda}$ and ${\rm Re}~a_{\Xi^0\Sigma^0}$. The latest
experimental measurements of all the WRHDs are collected in the Supplemental Material for the sake of easy reference. As shown in Table~\ref{tab:fitAmpb2tree}, there exist two sets of solutions with a reasonable $\chi^2/\mathrm{d.o.f.}$. The $\chi^2/{\mathrm{d.o.f.}}$ of Solution I is much smaller than that of Solution II. In addition,  $b_{\Xi^0\Lambda}^{(\rm 2,tree)}$ of Solution I is smaller, thus in  better agreement with the order-of magnitude estimates in naive dimensional analysis~\cite{Jenkins:1992ab,Manohar:1983md}. Therefore, we adopt the  $b_{\Xi^0\Lambda}^{(\rm 2,tree)}$ of Solution I to fix the contributions of other $b_{B_iB_f}^{(\rm 2,tree)}$.

\begin{table}[htpb]
\centering
\caption{\label{tab:fitAmpb2tree} Values of $b_{\Xi^0\Lambda}^{(\rm 2,tree)}$, ${\rm Re}~a_{\Xi^0\Lambda}$ and ${\rm Re}~a_{\Xi^0\Sigma^0}$  determined by fitting to the experimental branching fractions ${\cal B}$ and asymmetry parameters $\alpha_\gamma$ of the $\Xi^0\to\Sigma^0\gamma$ and $\Xi^0\to\Sigma^0\gamma$ decays.}
  \begin{tabular}{ccc}
\hline
\hline
 & Solution I & Solution II\\\cline{1-3}
\hline
$b_{\Xi^0\Lambda}^{(\rm 2,tree)}$ & $5.62(53)$ & $-8.34(48)$\\

${\rm Re}~a_{\Xi^0\Lambda}$ & $-9.56(34)$ & $3.89(45)$\\

${\rm Re}~a_{\Xi^0\Sigma^0}$ & $-32.22(64)$ & $32.50(61)$\\

$\chi^2/{\mathrm{d.o.f.}}$ & $0.04$ & $1.22$\\\cline{1-3}
\hline
\hline
\end{tabular}
\end{table}

Having obtained the contributions of $b_{B_iB_f}^{(\rm 2,tree)}$, $b_{B_iB_f}^{(\rm 2,loop)}$ and ${\rm Im}~a_{B_iB_f}^{(\rm 2,loop)}$, we can now  predict the real and imaginary parts of parity-violating amplitude $b$ and the imaginary part of parity-conserving amplitude $a$. The numerical results can be found in the Supplemental Material. In Fig.~\ref{Fig:PreObsLambton}, the predicted asymmetry parameters $\alpha$ in different approaches for the $\Lambda\to n\gamma$ decay are plotted  as a function of $\sqrt{|a|^2+|b|^2}$, which can be viewed as the branching fraction.  Interestingly, only the prediction of the EOMS B$\chi$PT agrees with the latest BESIII measurement~\cite{BESIII:2022rgl}. The different curves of the EOMS B$\chi{\rm PT}$ and HB $\chi{\rm PT}$ indicate that the asymmetry parameter $\alpha$ is particularly sensitive to amplitude $b$  because the real part of  amplitude $a$ is a free parameter while its imaginary part is constrained by unitarity. Furthermore, we note that the results of the pole model (PM I)~\cite{Gavela:1980bp} and $\chi$PT at  tree level~\cite{Borasoy:1999nt} are relatively closer to the new BESIII data. However,  the vector-dominance model~(VDM)~\cite{Zenczykowski:1991mx} and the other pole model~(PM II)~\cite{Nardulli:1987ub}  are disfavored by the BESIII data.

\begin{figure}[h!]
  \centering
  \includegraphics[width=0.9\linewidth]{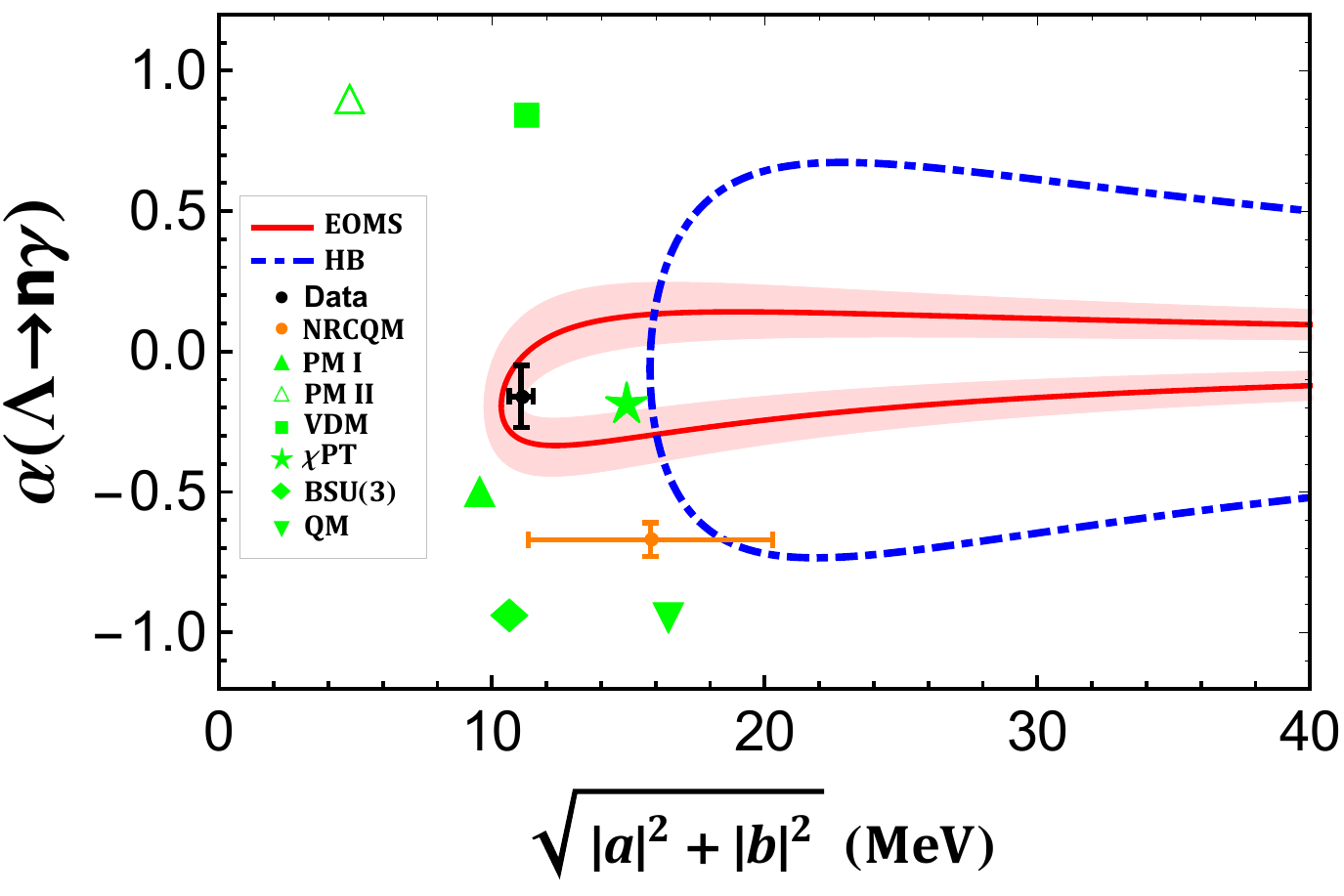}\\
  \caption{Asymmetry parameter $\alpha$ of  the $\Lambda\to n\gamma$ decay as a function of $\sqrt{|a|^2+|b|^2}$. The red solid line denotes the results of the EOMS B$\chi{\rm PT}$ and the blue dot-dashed line denotes the results of the HB $\chi{\rm PT}$. The band in lightred represents the uncertainties  originating mainly from amplitude $b$. The solid point in black with $xy$errorbars corresponds to the new BESIII data~\cite{BESIII:2022rgl} and that in orange is the prediction of the non-relativistic constituent quark model~(NRCQM)~\cite{Niu:2020aoz}. Other symbols in green stand for the results predicted in the pole models~(solid Delta: PM I~\cite{Gavela:1980bp} and hollow Delta: PM II~\cite{Nardulli:1987ub}), vector-dominance model~(solid square: VDM)~\cite{Zenczykowski:1991mx}, chiral perturbation theory at tree level~(solid star: $\chi{\rm PT}$)~\cite{Borasoy:1999nt}, broken SU(3)~(solid rhombus: BSU(3))~\cite{Zenczykowski:2005cs}, quark model~(solid nabla: QM)~\cite{Dubovik:2008zz}, respectively.}\label{Fig:PreObsLambton}
\end{figure}

\begin{figure}[htb!]
  \centering
  \includegraphics[width=0.9\linewidth]{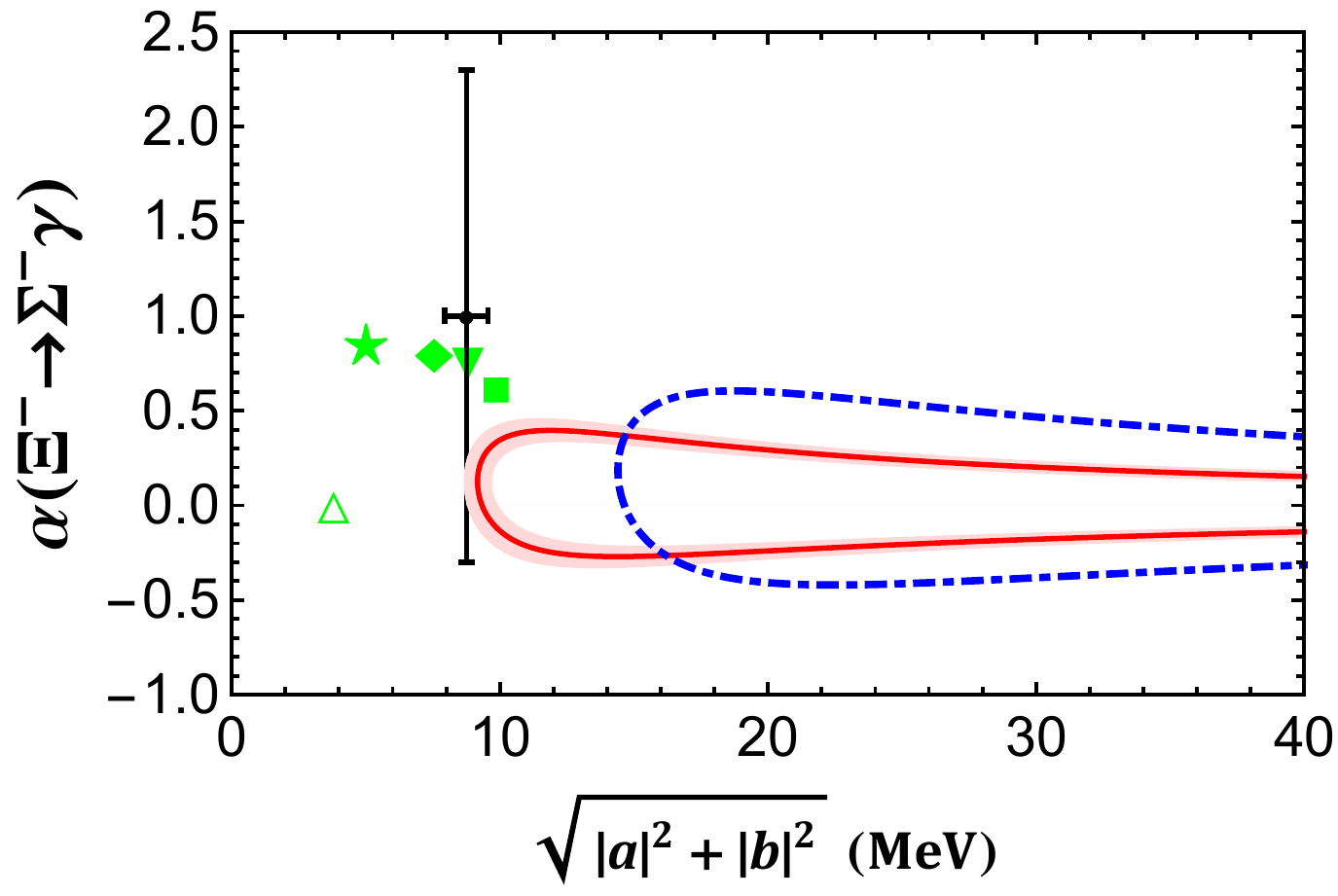}\\
  \includegraphics[width=0.9\linewidth]{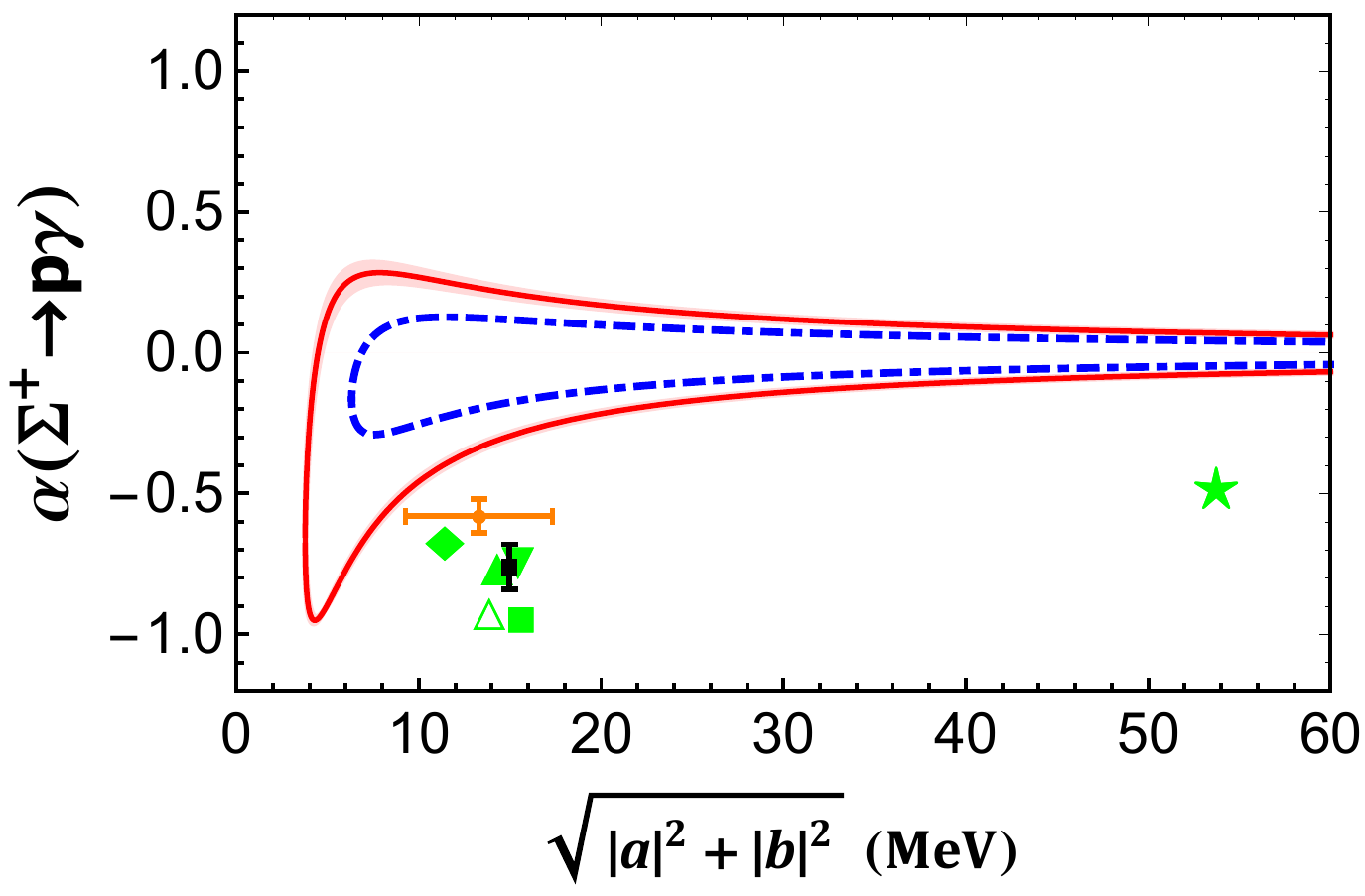}\\
  \includegraphics[width=0.9\linewidth]{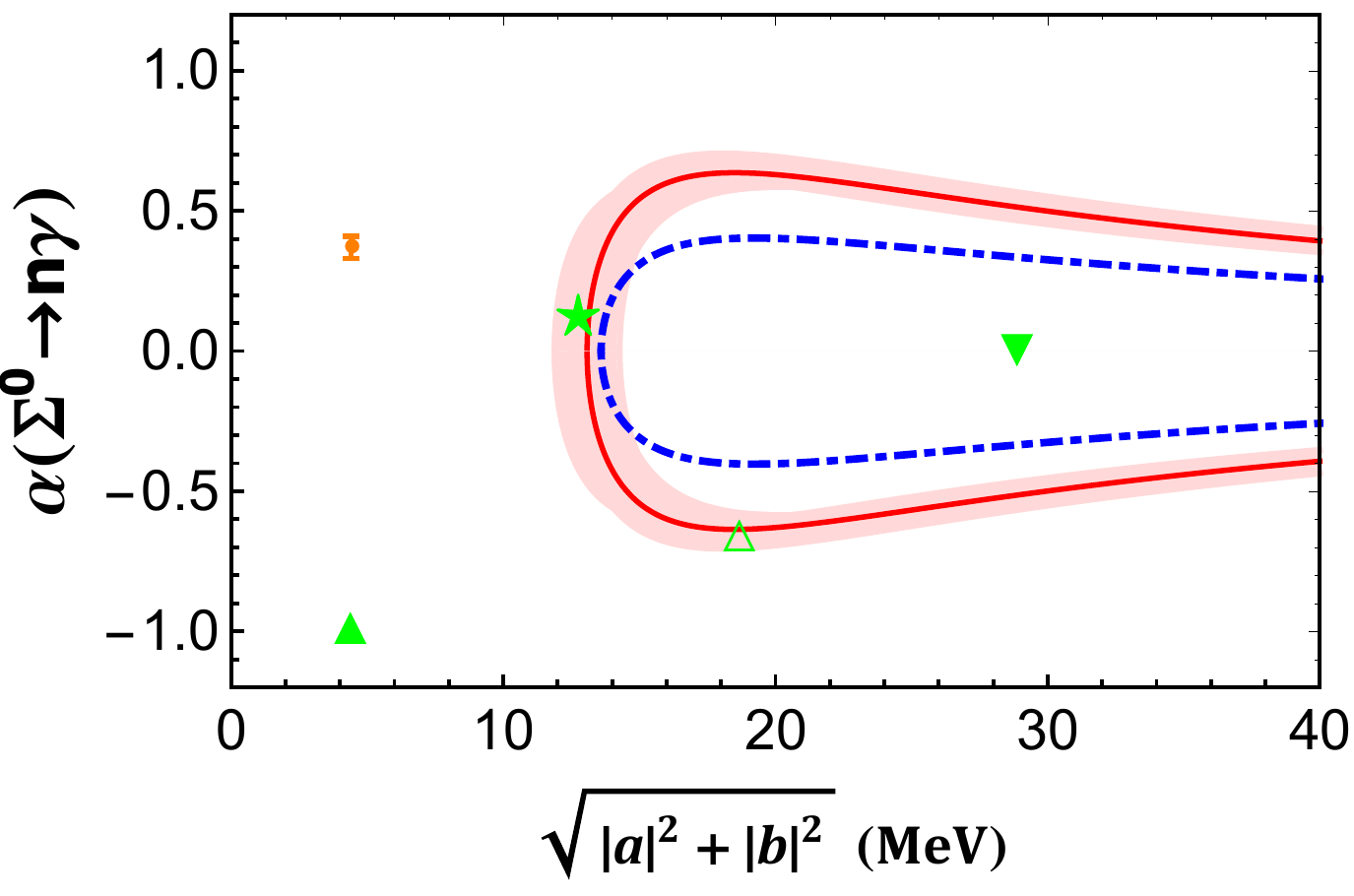}\\
  
  \caption{Same as Fig.~\ref{Fig:PreObsLambton} but for the $\Xi^-\to\Sigma^-\gamma$, $\Sigma^+\to p\gamma$, and $\Sigma^0\to n\gamma$  decays.}\label{Fig:PreObsOthers}
\end{figure}

Next, we show the predicted asymmetry parameters for the other three WRHDs as functions of $\sqrt{|a|^2+|b|^2}$  in Fig.~\ref{Fig:PreObsOthers}.
For the $\Xi^-\to\Sigma^-\gamma$ decay, our prediction agrees better with the experimental measurement than that of  the HB $\chi$PT, and the current PDG data disfavor the results of the pole model~(PM II)~\cite{Nardulli:1987ub} and tree-level~$\chi{\rm PT}$~\cite{Borasoy:1999nt}. For the $\Sigma^+\to p\gamma$ decay, the results predicted in all the $\chi$PT deviate from the PDG average but our prediction is closer. 
However, Hara’s theorem dictates that the asymmetry parameters for the $\Sigma^+\to p\gamma$ and $\Xi^-\to\Sigma^-\gamma$ decays should not be too large. With the constraints of the experimental branching fraction of $\Xi^-\to\Sigma^-\gamma$,  the allowed range of $\alpha_\gamma(\Xi^-\to\Sigma^-\gamma)$ is found to be $-0.18\sim 0.38$. Therefore, a more
precise measurement of the asymmetry parameter $\alpha_\gamma(\Xi^-\to\Sigma^-\gamma)$ is highly desirable in order to test Hara’s theorem and confirm the present experimental result. For the $\Sigma^0\to n\gamma$ decay, which has not been measured experimentally, we predict a lower limit for ${\cal B}(\Sigma^0\to n\gamma)$, which is $8.8\times10^{-13}$. This result contradicts the predictions of the pole model~(PM I)~\cite{Gavela:1980bp} and the non-relativistic constituent quark model~(NRCQM)~\cite{Niu:2020aoz}, which should be checked by future experiments.

{\it Summary:}
We showed that the latest precise measurement of the branching fraction and asymmetry parameter of the $\Lambda\to n\gamma$ by the BESIII Collaboration can be well explained  in covariant baryon chiral perturbation theory with the two relevant low energy constants $h_D$, $h_F$ and hyperon non-leptonic decay amplitudes determined by fitting to the latest experimental data on the $B_i\to B_f\pi$ decays and  the ${\cal O}(p^2)$ counter-term contributions determined by fitting to the $\Xi^0\to\Lambda\gamma$ and $\Xi^0\to\Sigma^0\gamma$ decays for the first time.    In addition, we predicted the asymmetry parameters for the $\Sigma^0\to n\gamma$ decay as a function of the real part of amplitude $a$, which, if measured in the future, can serve as a highly nontrivial check on the covariant baryon chiral perturbation theory and our understanding of the weak radiative hyperon decays.

We note that although the EOMS B$\chi$PT can describe well four of the five measured WRHDs, the predicted asymmetry parameter for  the $\Sigma^+\to p\gamma$ decay do not agree with the measurements. It has long been argued that the discrepancy  may  provide a window to new physics beyond the standard model~\cite{Cho:1993zb,Fujikawa:1993zu,Gabbiani:1996hi,He:1999ik,Tandean:1999mg}. Nonetheless, to seriously consider such proposals, we need to better understand the
non-perturbative strong interaction. In a series of works,  Borasoy and Holstein showed that some of the long-standing problems in hyperon non-leptonic decays and WRHDs could  be resolved by introducing explicitly the contribution of heavier resonances, such as the Delta, Roper, and $\Lambda(1405)$ multiplets~\cite{Borasoy:1998ku,Borasoy:1999md,Borasoy:1999nt}.  However, such studies need to be carefully reexamined.  For instance, the predicted decay parameters for $\Xi^0\rightarrow \Sigma^0\gamma$ and $\Xi^0\rightarrow \Lambda\gamma$ in Ref.~\cite{Borasoy:1999nt} are 0.15 and 0.46, respectively, in sharp conflict with the latest experimental data. We further note that the results of the present work are also important for studies of rare hyperon decays in searches for beyond the standard model physics~\cite{Geng:2021fog,He:2018yzu}.

{\it Acknowledgments:}
This work is supported in part by the National Natural Science Foundation of China under Grants No.11735003, No.11975041,  and No.11961141004. RXS acknowledges support from the National Natural Science Foundation of China under Grants No.12147145 and Project funded by China Postdoctoral Science Foundation No.2021M700343. JXL acknowledges support from the National Natural Science Foundation of China under Grant No.12105006 and China Postdoctoral Science Foundation under Grant No. 2021M690008.
\bibliography{WRHDs}

\clearpage

\section{Supplementary material for ``Weak radiative hyperon decays in covariant baryon chiral perturbation theory''}

In this supplemental material, we provide some details that are useful to understand the results presented in the main text.

\subsection{Contributions to the real parts of amplitudes $a$ and $b$ from the pole diagrams Figs.~(1a), (1b) and the direct photon emission diagram Fig.~(1c)}

From the relevant Lagrangian terms, one can easily obtain the real parts of amplitudes $a$ from the pole diagrams Figs.~(1a) and (1b), which are counted as of ${\cal O}(p)$, 
\begin{eqnarray}
a_{\Lambda n}^{\rm (1,tree)}&=&\frac{m_{\pi}^2F_\phi}{2m_B}\left[\frac{1}{\sqrt{3}}\left(h_D+3h_F\right)\frac{\mu_n^{(2)}-\mu_\Lambda^{(2)}}{m_\Lambda-m_n}\right.\nonumber\\
&-&\left.\left(h_D-h_F\right)\frac{\mu_{\Lambda\Sigma^0}^{(2)}}{m_{\Sigma^0}-m_n}\right],\nonumber\\
a_{\Sigma^+p}^{\rm (1,tree)}&=&\frac{m_{\pi}^2F_\phi}{2m_B}\left[-\sqrt{2}\left(h_D-h_F\right)\frac{\mu_p^{(2)}-\mu_{\Sigma^+}^{(2)}}{m_{\Sigma^+}-m_p}\right],\nonumber\\
a_{\Sigma^0n}^{\rm (1,tree)}&=&\frac{m_{\pi}^2F_\phi}{2m_B}\left[\left(h_D-h_F\right)\frac{\mu_n^{(2)}-\mu_{\Sigma^0}^{(2)}}{m_{\Sigma^0}-m_n}\right.\nonumber\\
&-&\left.\frac{1}{\sqrt{3}}\left(h_D+3h_F\right)\frac{\mu_{\Sigma^0\Lambda}^{(2)}}{m_\Lambda-m_n}\right],\nonumber\\
a_{\Xi^0\Lambda}^{\rm (1,tree)}&=&\frac{m_{\pi}^2F_\phi}{2m_B}\left[\frac{1}{\sqrt{3}}\left(h_D-3h_F\right)\frac{\mu_\Lambda^{(2)}-\mu_{\Xi^0}^{(2)}}{m_{\Xi^0}-m_{\Lambda}}\right.\nonumber\\
&+&\left.\left(h_D+h_F\right)\frac{\mu_{\Sigma^0\Lambda}^{(2)}}{m_{\Xi^0}-m_{\Sigma^0}}\right],\nonumber\\
a_{\Xi^0\Sigma^0}^{\rm (1,tree)}&=&\frac{m_{\pi}^2F_\phi}{2m_B}\left[\left(h_D+h_F\right)\frac{\mu_{\Sigma^0}^{(2)}-\mu_{\Xi^0}^{(2)}}{m_{\Xi^0}-m_{\Sigma^0}}\right.\nonumber\\
&+&\left.\frac{1}{\sqrt{3}}\left(h_D-3h_F\right)\frac{\mu_{\Lambda\Sigma^0}^{(2)}}{m_{\Xi^0}-m_{\Lambda}}\right],\nonumber\\
a_{\Xi^-\Sigma^-}^{\rm (1,tree)}&=&\frac{m_{\pi}^2F_\phi}{2m_B}\left[\sqrt{2}\left(h_D+h_F\right)
\frac{\mu_{\Xi^-}^{(2)}-\mu_{\Sigma^-}^{(2)}}{m_{\Xi^-}-m_{\Sigma^-}}\right],\label{Eq:OP1ampa}
\end{eqnarray}
where the number in the superscript of amplitude $a$  represents the chiral order and $\mu_B^{(2)}$ are the experimental baryon magnetic moments. We note that our results agree with those of Ref.~\cite{Jenkins:1992ab}.

The counter-term contributions at ${\cal O}(p^2)$ to the real parts of amplitudes $a$ and $b$ from the direct photon emission diagram Fig.~(1c) are given in Table~\ref{Tab:OP2ampsabtree}.  The counter-term contributions $b_{B_iB_f}^{(\rm 2,tree)}$ for each weak decay can be expressed in terms of the same LEC $C_\rho$, which implies the following relations:
\begin{eqnarray}
&&b_{\Xi^0\Sigma^0}^{(\rm 2,tree)}=\sqrt{3}b_{\Xi^0\Lambda}^{(\rm 2,tree)},~~~ b_{\Lambda n}^{(\rm 2,tree)}=-b_{\Xi^0\Lambda}^{(\rm 2,tree)},\nonumber\\
&&b_{\Sigma^0n}^{(\rm 2,tree)}=-\sqrt{3}b_{\Xi^0\Lambda}^{(\rm 2,tree)},~~~ b_{\Sigma^+p}^{(\rm 2,tree)}=0,~~~ b_{\Xi^-\Sigma^-}^{(\rm 2,tree)}=0.\label{Eq:countertermAmpb}
\end{eqnarray}
These results are those dictated by Hara’s theorem~\cite{Hara:1964zz}. In particular, we stress that the latter two terms imply a small asymmetry parameter.
\begin{table*}[htb!]
\renewcommand{\arraystretch}{1.5}
\centering
\caption{\label{Tab:OP2ampsabtree}Contributions to the real parts of amplitudes $a$ and $b$ from the direct photon emission diagram Fig.~(1c). The normalization $2(eG_F)^{-1}$ has been factored out.}
  \begin{tabular}{ccccccc}
\hline
\hline
  &~~~$\Lambda\to n\gamma$~~~ & ~~~$\Sigma^+\to p\gamma$~~~ & ~~~$\Sigma^0\to n\gamma$~~~ & ~~~$\Xi^0\to\Lambda\gamma$~~~ & ~~~$\Xi^0\to\Sigma^0\gamma$~~~ & ~~~$\Xi^-\to\Sigma^-\gamma$~~~\\
\hline
$a^{(\rm 2,tree)}$ & $\frac{2C_\alpha-C_\beta-C_\gamma+2C_\sigma}{3\sqrt{6}}$ & $\frac{2C_\beta-C_\gamma}{3}$ & $\frac{C_\beta+C_\gamma}{3\sqrt{2}}$ & $-\frac{C_\alpha-2C_\beta-2C_\gamma+C_\sigma}{3\sqrt{6}}$ & $\frac{C_\alpha+C_\sigma}{3\sqrt{2}}$ & $\frac{2C_\sigma-C_\alpha}{3}$\\

$b^{(\rm 2,tree)}$ & $-\frac{C_\rho}{\sqrt{6}}$ & $0$ & $-\frac{C_\rho}{\sqrt{2}}$ & $\frac{C_\rho}{\sqrt{6}}$ & $\frac{C_\rho}{\sqrt{2}}$ & $0$\\
\hline
\hline
\end{tabular}
\end{table*}

\subsection{Contributions to the imaginary parts of amplitudes $a$ from loop diagrams} 

Plugging the $P$-wave weak vertex  $-iG_Fm_\pi^2\bar{B}A_P\gamma_5B$ into the loop diagrams, one obtains the contributions to the imaginary parts of amplitudes $a$ which are of ${\cal O}(p^2)$:
\begin{eqnarray}
{\rm Im~a}_{\Lambda n}^{\rm (2,loop)}&=&\frac{m_{\pi}^2}{F_\pi}A_P(\Lambda\bar{p}\pi^+)g_A(p\bar{n}\pi^-)\nonumber\\
&\times&{\rm Im}\left[H_P^{(d)}(m_\Lambda,m_p,m_n,m_\pi)-H_P^{(f)}(m_\Lambda,m_p,m_n,m_\pi)\right],\nonumber\\
{\rm Im~a}_{\Sigma^+p}^{\rm (2,loop)}&=&-\frac{m_{\pi}^2}{F_\pi}A_P(\Sigma^+\bar{n}\pi^-)g_A(n\bar{p}\pi^+)\nonumber\\
&\times&{\rm Im}\left[H_P^{(d)}(m_{\Sigma^+},m_n,m_p,m_\pi)\right]\nonumber\\
&-&\frac{m_{\pi}^2}{F_\pi}A_P(\Sigma^+\bar{p}\pi^0)g_A(p\bar{p}\pi^0)\nonumber\\
&\times&{\rm Im}\left[H_P^{(f)}(m_{\Sigma^+},m_p,m_p,m_\pi)\right],\nonumber\\
{\rm Im~a}_{\Xi^-\Sigma^-}^{\rm (2,loop)}&=&\frac{m_{\pi}^2}{F_\pi}A_P(\Xi^-\bar{\Lambda}\pi^+)g_A(\Lambda\bar{\Sigma}^-\pi^-)\nonumber\\
&\times&{\rm Im}\left[H_P^{(d)}(m_{\Xi^-},m_\Lambda,m_{\Sigma^-},m_\pi)\right],\nonumber\\
{\rm Im~a}_{\Sigma^0n}^{\rm (2,loop)}&=&0,~~~{\rm Im~a}_{\Xi^0\Lambda}^{\rm (2,loop)}=0,~~~{\rm Im~a}_{\Xi^0\Sigma^0}^{\rm (2,loop)}=0,\label{Eq:loopampa}
\end{eqnarray}
where $A_P$ are the $P$-wave hyperon non-leptonic decay amplitudes, and $g_A$ is the strong interaction coupling determined by the LECs $D$ and $F$. The coefficients $A_p$, $g_A$ and loop functions $H_P^{(d,f)}(m_i,m_j,m_f,m_{\phi})$ can be found in the following.

The loop functions contributing to the imaginary parts of amplitudes $a$ read explicitly:
\begin{eqnarray}
H_P^{(d)}(m_i,m_j,m_f,m_\phi)&=&\int_0^{1}dx\int_0^{1-x}dy\left[2t_A\left(m_f^2x^2(x+y-1)\right.\right.\nonumber\\
&+&m_i^2xy(1-x)-m_im_jxy-m_im_fxy\nonumber\\
&+&\left.m_jm_fx(x+y)\right)+4t_B(3x-1)\nonumber\\
&-&\left.4d_Bx\right],\\
H_P^{(f)}(m_i,m_j,m_f,m_\phi)&=&\int_0^{1}dx\int_0^{1-x}dy\left[2t_A\left(m_f^2x(x+y-1)^2\right.\right.\nonumber\\
&+&m_i^2xy(1-x-y)-m_j^2y+m_im_jxy\nonumber\\
&+&\left.m_im_fxy-m_jm_f(x^2-2x+xy+1)\right)\nonumber\\
&+&4t_B(3x+3y-2)\nonumber\\
&-&\left.4d_B(x+y-1)\right].
\end{eqnarray}

The loop functions contributing to the real and imaginary parts of amplitudes $b$ are as follows:
\begin{eqnarray}
H_S^{(d)}(m_i,m_j,m_f,m_\phi)&=&\int_0^{1}dx\int_0^{1-x}dy\left[2t_A\left(m_f^2x^2(x+y-1)\right.\right.\nonumber\\
&+&m_i^2xy(1-x)+m_im_jxy+m_im_fxy\nonumber\\
&+&\left.m_jm_fx(x+y)\right)+4t_B(3x-1)\nonumber\\
&-&\left.4d_Bx\right],\\
H_S^{(e)}(m_i,m_j,m_f,m_\phi)&=&\int_0^{1}dx\int_0^{1-x}dy\left[2t_A\right.\nonumber\\
&\times&\left(m_f^2x(x-1)(x+y-1)-m_i^2x^2y\right.\nonumber\\
&+&m_im_jx(1-y)-m_im_fx(x+y-1)\nonumber\\
&-&\left.m_jm_fx(x+y-1)\right)+4t_B(3x-1)\nonumber\\
&-&\left.4d_Bx\right],\\
H_S^{(f)}(m_i,m_j,m_f,m_\phi)&=&\int_0^{1}dx\int_0^{1-x}dy\left[2t_A\left(m_f^2x(x+y-1)^2\right.\right.\nonumber\\
&+&m_i^2xy(1-x-y)-m_j^2y-m_im_jxy\nonumber\\
&-&\left.m_im_fxy-m_jm_f(x^2-2x+xy+1)\right)\nonumber\\
&+&4t_B(3x+3y-2)\nonumber\\
&-&\left.4d_B(x+y-1)\right],\\
H_S^{(g)}(m_i,m_j,m_f,m_\phi)&=&\int_0^{1}dx\int_0^{1-x}dy\left[2t_A\left(m_f^2xy(1-x-y)\right.\right.\nonumber\\
&+&m_i^2xy^2-m_j^2(1-x-y)\nonumber\\
&+&m_f(m_i+m_j)x(x+y-1)\nonumber\\
&+&\left.m_im_j(x+xy-1)\right)\nonumber\\
&+&\left.2t_B(2-6y)+4d_By\right],
\end{eqnarray}
with
\begin{eqnarray}
&&t_A=\frac{1}{32\pi^2}\frac{1}{\Delta},\nonumber\\
&&t_B=\frac{1}{64\pi^2}({\rm log}[\frac{\Delta}{\mu^2}]+1),\nonumber\\
&&d_B=-\frac{1}{64\pi^2}.
\end{eqnarray}
where  $\Delta=xy(m_f^2-m_i^2)+(1-x)(m_\phi^2-xm_f^2)+xm_j^2-i\epsilon$ for the loop functions of Feynman diagrams of Figs.~(1d) and (1e),  and $\Delta$ corresponding to Feynman diagrams of Figs.~(1f) and (1g) is $xy(m_f^2-m_i^2)+(1-x)(m_j^2-xm_f^2)+xm_\phi^2-i\epsilon$. In obtaining the above results, we have used the $\widetilde{\rm MS}$ scheme to regulate the loop functions. For the renormalization scale, we take $\mu=1$ GeV.

We have checked that the loop functions obtained in the EOMS scheme except for $H_P^{(d,f)}$ agree with those of Ref.~\cite{Neufeld:1992np}, because there the loop functions $H_P^{(d,f)}$ were calculated at ${\cal O}(p^3)$. We have verified that our ${\rm Im~a}$ for the loop functions $H_P^{(d,f)}$ at ${\cal O}(p^2)$ are consistent with those dictated by unitarity.

In Tables~\ref{Tab:AmpaCC} and \ref{Tab:AmpbCC}, we provide the coefficients appearing in the loop contributions to the real and imaginary parts of amplitudes $b$ and the imaginary parts of amplitudes $a$. 
\begin{table}[htb!]
\centering
\caption{\label{Tab:AmpaCC}Coefficients of the loop contributions to the imaginary parts of amplitudes $a$.}
  \begin{tabular}{cccc}
\hline
\hline
 &~~~~~~$A_p$~~~~~~ & ~~~~~~$g_A$~~~~~~\\
\hline
$\Lambda\to n\gamma$ & $A_P(\Lambda\bar{p}\pi^+)=11.61(19)$ & $g_A(p\bar{n}\pi^-)=-\frac{D+F}{\sqrt{2}}$\\
\hline
\multirow{2}{1.2cm}{$\Sigma^+\to p\gamma$} & $A_P(\Sigma^+\bar{n}\pi^-)=18.56(10)$ & $g_A(n\bar{p}\pi^+)=-\frac{D+F}{\sqrt{2}}$\\
& $A_P(\Sigma^+\bar{p}\pi^0)=12.94(40)$ & $g_A(p\bar{p}\pi^0)=-\frac{D+F}{2}$\\
\hline
$\Xi^-\to\Sigma^-\gamma$ & $A_P(\Xi^-\bar{\Lambda}\pi^+)=6.26(16)$ & $g_A(\Lambda\bar{\Sigma}^-\pi^-)=-\frac{D}{\sqrt{3}}$\\
\hline
\hline
\end{tabular}
\end{table}
\begin{table*}[htb!]
\renewcommand{\arraystretch}{1.8}
\centering
\caption{\label{Tab:AmpbCC}Coefficients of the loop contributions to the real and imaginary parts of amplitudes $b$.}
  \begin{tabular}{ccccc}
\hline
\hline
~~~Decay modes~~~ &~~~~~~$\xi_{B_j\phi}^{(d)}$~~~~~~ & ~~~~~~$\xi_{B_j\phi}^{(e)}$~~~~~~ & ~~~~~~$\xi_{B_j\phi}^{(f)}$~~~~~~& ~~~~~~$\xi_{B_j\phi}^{(g)}$~~~~~~\\
\hline
\multirow{3}{1.5cm}{$\Lambda\to n\gamma$} & $\xi_{\bar{\Sigma}^-K^-}^{(d)}=-\frac{h_D(D-F)}{\sqrt{3}}$ & $\xi_{\bar{\Sigma}^-\pi^-}^{(e)}=\frac{D(h_D-h_F)}{\sqrt{3}}$ & $\xi_{\bar{\Sigma}^-K^-}^{(f)}=\frac{h_D(D-F)}{\sqrt{3}}$ & $\xi_{\bar{\Sigma}^-\pi^-}^{(g)}=-\frac{D(h_D-h_F)}{\sqrt{3}}$\\

 & $\xi_{\bar{p}\pi^+}^{(d)}=\frac{(D+F)(h_D+3h_F)}{2\sqrt{3}}$ & $\xi_{\bar{p}K^+}^{(e)}=-\frac{(D+3F)(h_D+h_F)}{2\sqrt{3}}$ & $\xi_{\bar{p}\pi^+}^{(f)}=-\frac{(D+F)(h_D+3h_F)}{2\sqrt{3}}$ & $\xi_{\bar{p}K^+}^{(g)}=\frac{(D+3F)(h_D+h_F)}{2\sqrt{3}}$\\

 & $\cdots$ & $\xi_{\bar{\Sigma}^+\pi^+}^{(e)}=0.06\cdot\frac{D}{\sqrt{3}}$ & $\cdots$ & $\xi_{\bar{\Sigma}^+\pi^+}^{(g)}=-0.06\cdot\frac{D}{\sqrt{3}}$\\
\hline
\multirow{3}{1.5cm}{$\Sigma^+\to p\gamma$} & $\xi_{\bar{\Lambda} K^-}^{(d)}=\frac{h_D(D+3F)}{3\sqrt{2}}$ & $\xi_{\bar{\Lambda}\pi^-}^{(e)}=-\frac{D(h_D+3h_F)}{3\sqrt{2}}$ & $\xi_{\bar{\Sigma}^+K^0}^{(f)}=-\frac{(h_D-h_F)(D-F)}{\sqrt{2}}$ & $\xi_{\bar{p}K^0}^{(g)}=\frac{(h_D-h_F)(D-F)}{\sqrt{2}}$\\

 & $\xi_{\bar{\Sigma}^0 K^-}^{(d)}=\frac{h_F(D-F)}{\sqrt{2}}$ & $\xi_{\bar{\Sigma}^0\pi^-}^{(e)}=-\frac{F(h_D-h_F)}{\sqrt{2}}$ & $\xi_{\bar{p}\pi^0}^{(f)}=-\frac{(h_D-h_F)(D+F)}{2\sqrt{2}}$ & $\xi_{\bar{p}\pi^0}^{(g)}=-\frac{F(h_D-h_F)}{\sqrt{2}}$\\

 & $\xi_{\bar{n}\pi^-}^{(d)}=-0.06\cdot\frac{D+F}{\sqrt{2}}$ & $\cdots$ & $\xi_{\bar{p}\eta}^{(f)}=-\frac{(h_D-h_F)(D-3F)}{2\sqrt{2}}$ & $\xi_{\bar{p}\eta}^{(g)}=\frac{D(h_D-h_F)}{\sqrt{2}}$\\
\hline
\multirow{3}{1.5cm}{$\Sigma^0\to n\gamma$} & $\xi_{\bar{\Sigma}^-K^-}^{(d)}=-h_F(D-F)$ & $\xi_{\bar{\Sigma}^-\pi^-}^{(e)}=F(h_D-h_F)$ & $\xi_{\bar{\Sigma}^-K^-}^{(f)}=h_F(D-F)$ & $\xi_{\bar{\Sigma}^-\pi^-}^{(g)}=-F(h_D-h_F)$\\

 & $\xi_{\bar{p}\pi^+}^{(d)}=-\frac{(h_D-h_F)(D+F)}{2}$ & $\xi_{\bar{p}K^+}^{(e)}=\frac{(h_D+h_F)(D-F)}{2}$ & $\xi_{\bar{p}\pi^+}^{(f)}=\frac{(h_D-h_F)(D+F)}{2}$ & $\xi_{\bar{p}K^+}^{(g)}=-\frac{(h_D+h_F)(D-F)}{2}$\\

 & $\cdots$ & $\xi_{\bar{\Sigma}^+\pi^+}^{(e)}=-0.06F$ & $\cdots$ & $\xi_{\bar{\Sigma}^+\pi^+}^{(g)}=0.06F$\\
\hline
\multirow{2}{1.5cm}{$\Xi^0\to\Lambda\gamma$} &  $\xi_{\bar{\Sigma}^+\pi^+}^{(d)}=-\frac{D(h_D+h_F)}{\sqrt{3}}$ & $\xi_{\bar{\Sigma}^+K^+}^{(e)}=\frac{h_D(D+F)}{\sqrt{3}}$ &  $\xi_{\bar{\Sigma}^+\pi^+}^{(f)}=\frac{D(h_D+h_F)}{\sqrt{3}}$ & $\xi_{\bar{\Sigma}^+K^+}^{(g)}=-\frac{h_D(D+F)}{\sqrt{3}}$\\

 & $\xi_{\bar{\Xi}^-K^-}^{(d)}=\frac{(h_D-h_F)(D-3F)}{2\sqrt{3}}$ & $\xi_{\bar{\Xi}^-\pi^-}^{(e)}=-\frac{(h_D-3h_F)(D-F)}{2\sqrt{3}}$ & $\xi_{\bar{\Xi}^-K^-}^{(f)}=-\frac{(h_D-h_F)(D-3F)}{2\sqrt{3}}$ & $\xi_{\bar{\Xi}^-\pi^-}^{(g)}=\frac{(h_D-3h_F)(D-F)}{2\sqrt{3}}$\\
\hline
\multirow{2}{1.5cm}{$\Xi^0\to\Sigma^0\gamma$} & $\xi_{\bar{\Sigma}^+\pi^+}^{(d)}=F(h_D+h_F)$ & $\xi_{\bar{\Sigma}^+K^+}^{(e)}=-h_F(D+F)$ & $\xi_{\bar{\Sigma}^+\pi^+}^{(f)}=-F(h_D+h_F)$ & $\xi_{\bar{\Sigma}^+K^+}^{(g)}=h_F(D+F)$\\

 & $\xi_{\bar{\Xi}^-K^-}^{(d)}=-\frac{(h_D-h_F)(D+F)}{2}$ & $\xi_{\bar{\Xi}^-\pi^-}^{(e)}=\frac{(h_D+h_F)(D-F)}{2}$ & $\xi_{\bar{\Xi}^-K^-}^{(f)}=\frac{(h_D-h_F)(D+F)}{2}$ & $\xi_{\bar{\Xi}^-\pi^-}^{(g)}=-\frac{(h_D+h_F)(D-F)}{2}$\\
\hline
\multirow{3}{1.5cm}{$\Xi^-\to\Sigma^-\gamma$} & $\xi_{\bar{\Lambda}\pi^+}^{(d)}=\frac{D(h_D-3h_F)}{3\sqrt{2}}$ & $\xi_{\bar{\Lambda}K^+}^{(e)}=-\frac{h_D(D-3F)}{3\sqrt{2}}$ & $\xi_{\bar{\Xi}^-\pi^0}^{(f)}=-\frac{F(h_D+h_F)}{\sqrt{2}}$ & $\xi_{\bar{\Xi}^-\pi^0}^{(g)}=\frac{(h_D+h_F)(D-F)}{2\sqrt{2}}$\\

 & $\xi_{\bar{\Sigma}^0\pi^+}^{(d)}=-\frac{F(h_D+h_F)}{\sqrt{2}}$ & $\xi_{\bar{\Sigma}^0K^+}^{(e)}=\frac{h_F(D+F)}{\sqrt{2}}$ & $\xi_{\bar{\Xi}^-\eta}^{(f)}=-\frac{D(h_D+h_F)}{\sqrt{2}}$ & $\xi_{\bar{\Xi}^-\eta}^{(g)}=\frac{(h_D+h_F)(D+3F)}{2\sqrt{2}}$\\
 
 & $\cdots$ & $\cdots$ & $\xi_{\bar{\Xi}^-K^0}^{(f)}=-\frac{(h_D+h_F)(D+F)}{\sqrt{2}}$ & $\xi_{\bar{\Xi}^-K^0}^{(g)}=\frac{(h_D+h_F)(D+F)}{\sqrt{2}}$\\
\hline
\hline
\end{tabular}
\end{table*}

\subsection{Values of LECs $h_D$, $h_F$ and hyperon non-leptonic decay amplitudes}

As mentioned in the main text, one needs to know the values of LECs $h_D$, $h_F$ and the hyperon non-leptonic decay amplitudes in order to predict the parity-conserving and parity-violating  amplitudes $a$ and $b$. These relevant values were determined by fitting to the 1992 PDG data for hyperon non-leptonic decays~\cite{Jenkins:1991bt}. However, the most recent experimental data~\cite{ParticleDataGroup:2020ssz,BESIII:2018cnd,Ablikim:2022ick,BESIII:2021ypr,Ireland:2019uja} have changed significantly compared to the previous PDG averages. Especially,
the latest BESIII measurement of the $\Lambda$ baryon decay parameter $\alpha_\pi(\Lambda\rightarrow p\pi^-)$~\cite{BESIII:2018cnd,Ablikim:2022ick,BESIII:2021ypr,Ireland:2019uja} is larger than the earlier PDG average by more than $5\sigma$. We note that the BESIII Collaboration in 2019~\cite{BESIII:2018cnd} and 2022~\cite{Ablikim:2022ick}
used the same method to measure  $\alpha_\pi(\Lambda\rightarrow p\pi^-)$ and obtained similar results, but the uncertainty of the BESIII measurement in 2022~\cite{Ablikim:2022ick} is smaller. As a result, we have taken the value of $\alpha_\pi(\Lambda\rightarrow p\pi^-)$ as an average obtained in three different measurements~\cite{Ablikim:2022ick,BESIII:2021ypr,Ireland:2019uja}(see Table~\ref{Tab:SPampexp}). The above discussion shows that the results obtained in Ref~\cite{Jenkins:1991bt} are out of date. Therefore, we need to update the study of hyperon non-leptonic decays and check how the latest data affect the values of the LECs $h_D$, $h_F$ and hyperon non-leptonic decay amplitudes.

The hyperon non-leptonic decay amplitudes for the octet-to-octet transitions have the following form
\begin{eqnarray}
{\cal M}(B_i\to B_f\pi)=iG_Fm_\pi^2\bar{B}_f\left(A_S-A_P\gamma_5\right)B_i,\label{Eq:NonLepHDs}
\end{eqnarray}
where the dimensionless $A_S$ and $A_P$ denote $S$- and $P$-wave amplitudes, respectively. One should note that $A_P$ is of order $m_B$ despite that $\bar{B}_f\gamma_5B_i$ is suppressed by $1/m_B$. As a result, the $P$-wave contribution to the total amplitude is of the same order as the $S$-wave contribution, both of which are ${\cal O}(p^0)$. This point has  been noted in Ref.~\cite{Jenkins:1991bt}. For the $B_i\to B_f\pi$ decays, the corresponding partial widths and baryon decay parameters are, respectively:
\begin{eqnarray}
&&\Gamma(B_i\to B_f\pi)=\frac{\left(G_Fm_\pi^2\right)^2}{8\pi m_i^2}\left|\vec{q}\right|\left\{[(m_i+m_f)^2-m_\pi^2]\left|s\right|^2\right.\nonumber\\
&&\left.\qquad\qquad\qquad+[(m_i-m_f)^2-m_\pi^2]
\left|p\cdot\frac{(E_f+m_f)}{\left|\vec{q}\right|}\right|^2\right\},\nonumber\\
&&\alpha_\pi=\frac{2{\rm Re}\left(s\cdot p\right)}{\left|s\right|^2+\left|p\right|^2},~
\beta_\pi=\frac{2{\rm Im}\left(s\cdot p\right)}{\left|s\right|^2+\left|p\right|^2},
~\gamma_\pi=\frac{\left|s\right|^2+\left|p\right|^2}{\left|s\right|^2+\left|p\right|^2},
\label{Eq:NonLepHDsObs}
\end{eqnarray}
where $E_f$ and $|\vec{q}|$ are the energy and 3-momentum of the final baryon,  $s=A_S$ and $p=A_P|\vec{q}|/(E_f+m_f)$ with $|\vec{q}|=\frac{1}{2m_i}\lambda^{1/2}(m_i^2,m_f^2,m_\pi^2)$, and $\lambda(a,b,c)=a^2+b^2+c^2-2(ab+ac+bc)$.
\begin{table*}[htb!]
\caption{\label{Tab:SPampexp}Experimental  $S$- and $P$-wave hyperon non-leptonic decay amplitudes extracted from the most recent pdgLive~\cite{ParticleDataGroup:2020ssz}, BESIII measurements~\cite{Ablikim:2022ick,BESIII:2021ypr} and CLAS data~\cite{Ireland:2019uja}.}
\centering
  \begin{tabular}{ccccccccc}
\hline
\hline
\multirow{2}{2cm}{Decay modes} & ~~\multirow{2}{1cm}{${\cal B}$~\cite{ParticleDataGroup:2020ssz}} & ~~\multirow{2}{2cm}{$\alpha_\pi$~\cite{ParticleDataGroup:2020ssz,Ablikim:2022ick,BESIII:2021ypr,Ireland:2019uja}} & ~~~\multirow{2}{2cm}{$\phi_\pi~(^\circ)$~\cite{ParticleDataGroup:2020ssz,BESIII:2021ypr}}~~~ & \multicolumn{2}{c}{$s=A_S^{(\rm Expt)}$} &~~~& \multicolumn{2}{c}{$p=A_P^{(\rm Expt)}|\vec{q}|/(E_f+m_f)$}\\
\cline{5-6}\cline{8-9}
& & & &~~~{\rm This work}~~~ & ~\cite{Jenkins:1991bt} & & ~~~{\rm This work}~~~ & ~\cite{Jenkins:1991bt}\\
\hline
$\Sigma^+\to n\pi^+$ & $0.4831(30)$ & $0.068(13)$ & $167(20)$ & $0.06(1)$ & $0.06(1)$ & &$1.81(1)$ & $1.81(1)$\\

$\Sigma^-\to n\pi^-$ & $0.99848(5)$ & $-0.068(8)$ & $10(15)$ & $1.88(1)$ & $1.88(1)$ & &$-0.06(1)$ & $-0.06(1)$\\

$\Lambda\to p\pi^-$ & $0.639(5)$ & $0.7462(88)$ & $-6.5(35)$ & $1.38(1)$ & $1.42(1)$ & &$0.62(1)$ & $0.52(2)$\\

$\Xi^-\to\Lambda\pi^-$ & $0.99887(35)$ & $-0.376(8)$ & $0.6(12)$ & $-1.99(1)$ & $-1.98(1)$ & &$0.39(1)$ & $0.48(2)$\\

$\Sigma^+\to p\pi^0$ & $0.5157(30)$ & $-0.982(14)$ & $36(34)$ & $-1.50(3)$ & $-1.43(5)$ & &$1.29(4)$ & $1.17(7)$\\

$\Lambda\to n\pi^0$ & $0.358(5)$ & $0.74(5)$ & $\cdots$ & $-1.09(2)$ & $-1.04(1)$ & &$-0.48(4)$ & $-0.39(4)$\\

$\Xi^0\to\Lambda\pi^0$ & $0.99524(12)$ & $-0.356(11)$ & $21(12)$ & $1.62(10)$ & $1.52(2)$ & &$-0.30(10)$ & $-0.33(2)$\\
\hline
\hline
\end{tabular}
\end{table*}

By means of isospin symmetry, the Lee-Sugawara relations~\cite{Sugawara:1964zz,Lee:1964zzc} and the criterion that $A_S(\Lambda\to p\pi^-)$ is conventionally positive, the $S$- and $P$-wave contributions to the total decay amplitudes  can be determined solely by fitting to the measured branching fractions, $\alpha_\pi$ and $\gamma_\pi$ for hyperon non-leptonic decays~\cite{ParticleDataGroup:2020ssz,Ablikim:2022ick,BESIII:2021ypr,Ireland:2019uja}. The relevant results are tabulated in Table~\ref{Tab:SPampexp}. Here the experimental values of baryon decay parameters $\gamma_\pi$ for each decay mode are extracted from  $\alpha_\pi$ and  $\phi_\pi$ collected in Table~\ref{Tab:SPampexp}. The three quantities are related:
\begin{eqnarray}
\gamma_\pi=\sqrt{1-\alpha_\pi^2}\cos\left(\phi_\pi\right).
\end{eqnarray}
Comparing our results with those of Ref.~\cite{Jenkins:1991bt}, we find that the experimental  $S$-wave amplitudes for the $\Lambda\to p\pi^-$ and $\Lambda\to n\pi^0$ decays are slightly different, which would affect mildly the values of the two LECs $h_D$ and $h_F$. In contrast, the $P$-wave amplitudes differ a lot. Especially,  for the $\Lambda\to p\pi^-$ and $\Xi^-\to\Lambda^-\gamma$ decays, the deviations from the previous results~\cite{Jenkins:1991bt} are about $4\sim5\sigma$. From Eq.~(\ref{Eq:loopampa}), we can clearly see that a large change in the $P$-wave amplitudes will greatly affect the imaginary parts of the parity-conserving amplitude $a$.
\begin{table}[h!]
\caption{\label{Tab:TwoLECsfit}LECs $h_D$ and $h_F$ determined by fitting to the $S$-wave hyperon non-leptonic decay amplitudes.}
\centering
  \begin{tabular}{cccc}
\hline
\hline
~~~Decay modes~~~ & ~~~$A_S^{\rm th}$~~~ & ~~~$A_S^{\rm Expt}$~~~\\
\hline
$\Sigma^+\to n\pi^+$ & $0$ & $0.06(1)$\\

$\Sigma^-\to n\pi^-$ & $-h_D+h_F$ & $1.88(1)$\\

$\Lambda\to p\pi^-$ & $\frac{1}{\sqrt{6}}(h_D+3h_F)$ & $1.38(1)$\\

$\Xi^-\to\Lambda\pi^-$ & $\frac{1}{\sqrt{6}}(h_D-3h_F)$ & $-1.99(1)$\\

$\Sigma^+\to p\pi^0$ & $\frac{1}{\sqrt{2}}(h_D-h_F)$ & $-1.50(3)$\\

$\Lambda\to n\pi^0$ & $-\frac{1}{2\sqrt{3}}(h_D+3h_F)$ & $-1.09(2)$\\

$\Xi^0\to\Lambda\pi^0$ & $-\frac{1}{2\sqrt{3}}(h_D-3h_F)$ & $1.62(10)$\\
\hline
$\chi^2/\mathrm{d.o.f.}=0.24$ & $h_D=-0.61(24)$ & $h_F=1.42(14)$\\
\hline
\hline
\end{tabular}
\end{table}

As pointed out in Ref.~\cite{Jenkins:1991bt}, if the two LECs $h_D$ and $h_F$ can describe well the experimental $S$-wave amplitudes, they reproduce very poorly the $P$-wave amplitudes, which is the so-called $S/P$ puzzle. In this work, we donot discuss this long-standing problem. As a result, we only updated the values of $h_D$ and $h_F$ by  fitting to the experimental  $S$-wave amplitudes for hyperon non-leptonic decays. We note that the experimental  $S$-wave amplitudes are rather precisely measured. Therefore, in our least-squares fit, an absolute uncertainty of 0.3 is added to each $S$-wave amplitude in order to match the theoretical predictions with the experimental data at $1\sigma$ confidence level. In Table~\ref{Tab:TwoLECsfit}, we present the tree-level formulae for the $S$-wave amplitudes derived from Lagrangian ${\cal L}_{\Delta S=1}^{(0)}$ and find that the central values of LECs $h_D$ and $h_F$ deviate slightly from those of Ref.~\cite{Jenkins:1991bt}, i.e., $h_D=-0.58(21)$ and $h_F=1.40(12)$. Nevertheless, our results are consistent with Ref.~\cite{Jenkins:1991bt} within $1\sigma$ confidence intervals.

\subsection{Predictions for parity-conserving $a$ and parity-violating $b$ amplitudes}

As stated in the main text and Supplemental Material, the counter-term contributions $b^{\rm (2,tree)}$ to the real parts of amplitudes $b$ at ${\cal O}(p^2)$ are determined by the experimental branching fractions and asymmetry parameters of $\Xi^0\to\Lambda\gamma$ and $\Xi^0\to\Sigma^0\gamma$. The amplitudes ${\rm Im}~a_{B_iB_f}^{(\rm 2,loop)}$ and $b_{B_iB_f}^{(\rm 2,loop)}$ from loop contributions at ${\cal O}(p^2)$ can be obtained by using the four known LECs $h_D$, $h_F$, $D$, $F$ and hyperon non-leptonic decay amplitudes from Table~\ref{Tab:SPampexp}.

In Tables~\ref{Tab:AmpbPred} and \ref{Tab:AmpaPred}, we decompose the contributions to amplitudes $b$ of each decay mode into those from the tree and loop diagrams at ${\cal O}(p^2)$, and compare the  amplitudes $a$ and $b$ obtained in the present work with those obtained in the HB $\chi{\rm PT}$ of Ref.~\cite{Jenkins:1992ab} and B$\chi$PT of~\cite{Neufeld:1992np}. The uncertainties  stem from the LECs $h_D$, $h_F$ and experimental hyperon non-leptonic decay amplitudes. Comared with those of the HB $\chi{\rm PT}$~\cite{Jenkins:1992ab} and B$\chi$PT~\cite{Neufeld:1992np}, our results for the first time included the contributions of the  ${\cal O}(p^2)$ counter-terms. It should be stressed that in our analysis, the B$\chi$PT results~\cite{Neufeld:1992np} can be taken as the ones obtained from ${\cal O}(p^2)$ loop diagrams with the EOMS scheme for the  reasons given below. In addition, as pointed out in Refs.~\cite{Neufeld:1992np,He:2005yn}, some sizable difference between the EOMS B$\chi{\rm PT}$ and HB $\chi{\rm PT}$ results for loop-level amplitudes $a$ and $b$  arise mainly from relativistic corrections.
\begin{table*}[htb!]
\centering
\caption{\label{Tab:AmpbPred} Decomposition of the contributions to the parity-violating amplitudes $b$  (in units of MeV).}
  \begin{tabular}{cccccccc}
\hline
\hline
  & \multicolumn{3}{c}{EOMS B$\chi{\rm PT}$} & ~ & B$\chi{\rm PT}$~\cite{Neufeld:1992np} & ~ & HB $\chi{\rm PT}$~\cite{Jenkins:1992ab}\\\cline{2-4}\cline{6-6}\cline{8-8}
Decay modes &~~$b^{\rm (2,tree)}$ & ~~$b^{\rm (2,loop)}$ & ~~$b^{\rm (2,tot)}$ & & ~~$b^{\rm (2,tot)}=b^{\rm (2,loop)}$ & &~~$b^{\rm (2,tot)}=b^{\rm (2,loop)}$\\
\hline
$\Lambda\to n\gamma$ & -5.62(53) &$7.87(73)+10.04(81)i$  & $2.25(90)+10.04(81)i$ & & $7.22+9.59i$ & & $11.11+11.21i$\\

$\Sigma^+\to p\gamma$ & $0$  & $-1.96(11)-1.75(12)i$ & $-1.96(11)-1.75(12)i$ & & $-1.42-1.42i$ & & $-1.21-0.53i$\\

$\Sigma^0\to n\gamma$ & $-9.73(92)$ &$1.41(11)+10.09(78)i$ & $-8.32(93)+10.09(78)i$ & & $1.28+9.46i$ & & $5.48+12.44i$\\

$\Xi^0\to\Lambda\gamma$ & $5.62(53)$ & $-1.60(48)$ & $4.02(72)$ & &$-1.64$ & & $-2.47$\\

$\Xi^0\to\Sigma^0\gamma$ & $9.73(92)$ & $2.91(67)$ & $12.64(114)$ & & $2.56$ & & $2.52$\\

$\Xi^-\to \Sigma^-\gamma$ & $0$ & $-3.00(29)-8.64(54)i$ & $-3.00(29)-8.64(54)i$ & & $-2.60-8.00i$ & &$-7.26-12.34i$\\
\hline
\hline
\end{tabular}
\end{table*}

\begin{table}[htb!]
\centering
\caption{\label{Tab:AmpaPred}Imaginary parts of the loop contributions to the parity-conserving amplitudes $a$  at ${\cal O}(p^2)$ (in units of MeV).}
  \begin{tabular}{cccccc}
\hline
\hline
 & ~~EOMS B$\chi{\rm PT}$ &~& ~~B$\chi{\rm PT}$~\cite{Neufeld:1992np} &~& ~~HB $\chi{\rm PT}$~\cite{Jenkins:1992ab}\\\cline{2-2}\cline{4-4}\cline{6-6}
Decay modes &${\rm Im}~a^{\rm (2,loop)}$ &~& ${\rm Im}~a^{\rm (unitarity)}$ &~& ${\rm Im}~a^{\rm (unitarity)}$\\\cline{1-4}
\hline
$\Lambda\to n\gamma$ & $-1.01(2)$ & & $-0.82$ & &$-0.68$\\

$\Sigma^+\to p\gamma$ & $2.70(4)$  & &$2.60$ & &$6.18$\\

$\Xi^-\to \Sigma^-\gamma$ & $-0.57(1)$  & & $-0.64$ & &$-1.55$\\\cline{1-4}
\hline
\hline
\end{tabular}
\end{table}

\subsection{Comparison between the EOMS B$\chi$PT and B$\chi$PT results}

In this section, we highlight the differences between the EOMS B$\chi$PT and B$\chi$PT results of Ref.~\cite{Neufeld:1992np}.
Our results for amplitudes $b$ contain the contributions of counter-terms and of the experimentally measured $S$-wave amplitude for the hyperon non-leptonic decay $\Sigma^+\to n\pi^+$, where $A_S(\Sigma^+\to n\pi^+)$ vanishes in the SU(3) limit. In the absence of $b^{\rm (2,tree)}$ contributions, we note that there is a slight difference between the EOMS B$\chi{\rm PT}$ and B$\chi{\rm PT}$ results~\cite{Neufeld:1992np}. This is mainly because in Ref.~\cite{Neufeld:1992np} $F_\phi=F_\pi$, $D=0.756(11)$ and $A_S(\Sigma^+\to n\pi^+)=0$, while we take $F_K=1.22F_\pi$, $F_\eta=1.3F_\pi$, $D=0.793(18)$ and $A_S(\Sigma^+\to n\pi^+)=0.06(1)$. For the imaginary parts of amplitudes $a$, our results calculated from the ${\cal O}(p^2)$ loop diagrams  are very close to those obtained by unitarity in Ref.~\cite{Neufeld:1992np} with the exception of $\Lambda\to n\gamma$. The difference of ${\rm Im~a}(\Lambda\to n\gamma)$ originates from the fact that the updated $P$-wave amplitude for the $\Lambda\to p\pi^-$ decay used in our work is different from that used in Ref.~\cite{Neufeld:1992np} by $5\sigma$. With the same input values, we can recover the results of Ref.~\cite{Neufeld:1992np}.

\subsection{Experimental branching fractions ${\cal B}$ and asymmetry parameters $\alpha_\gamma$ for the WRHDs}
\begin{table}[htb!]
\centering
\caption{\label{tab:Expdata} Branching fractions ${\cal B}$ and asymmetry parameters $\alpha_\gamma$ for the WRHDs~\cite{E761:1993unn,ParticleDataGroup:2020ssz,BESIII:2022rgl}.}
  \begin{tabular}{ccc}
\hline
\hline
 Decay modes & ${\cal B}\times10^{-3}$ & $\alpha_\gamma$\\\cline{1-3}
\hline
$\Lambda\to n\gamma$ & $0.832(38)(54)$ & $-0.16(10)(50)$\\

$\Sigma^+\to p\gamma$ & $1.23(5)$ & $-0.76(8)$\\

$\Sigma^0\to n\gamma$ & $\cdots$ & $\cdots$\\

$\Xi^0\to\Lambda\gamma$ & $1.17(7)$ & $-0.704(19)(64)$\\

$\Xi^0\to\Sigma^0\gamma$ & $3.33(10)$ & $-0.69(6)$\\

$\Xi^-\to \Sigma^-\gamma$ & $0.127(23)$ & $1.0(13)$\\\cline{1-3}
\hline
\hline
\end{tabular}
\end{table}

For easy reference, we collect all the latest experimental branching fractions and asymmetry parameters  in Table~\ref{tab:Expdata}.

\subsection{Asymmetry parameters of the $\Xi^0\to\Lambda\gamma$ and $\Xi^0\to\Sigma^0\gamma$ decays as  functions of $\sqrt{a^2+b^2}$}

\begin{figure*}[htb!]
  \centering
  \includegraphics[width=0.45\linewidth]{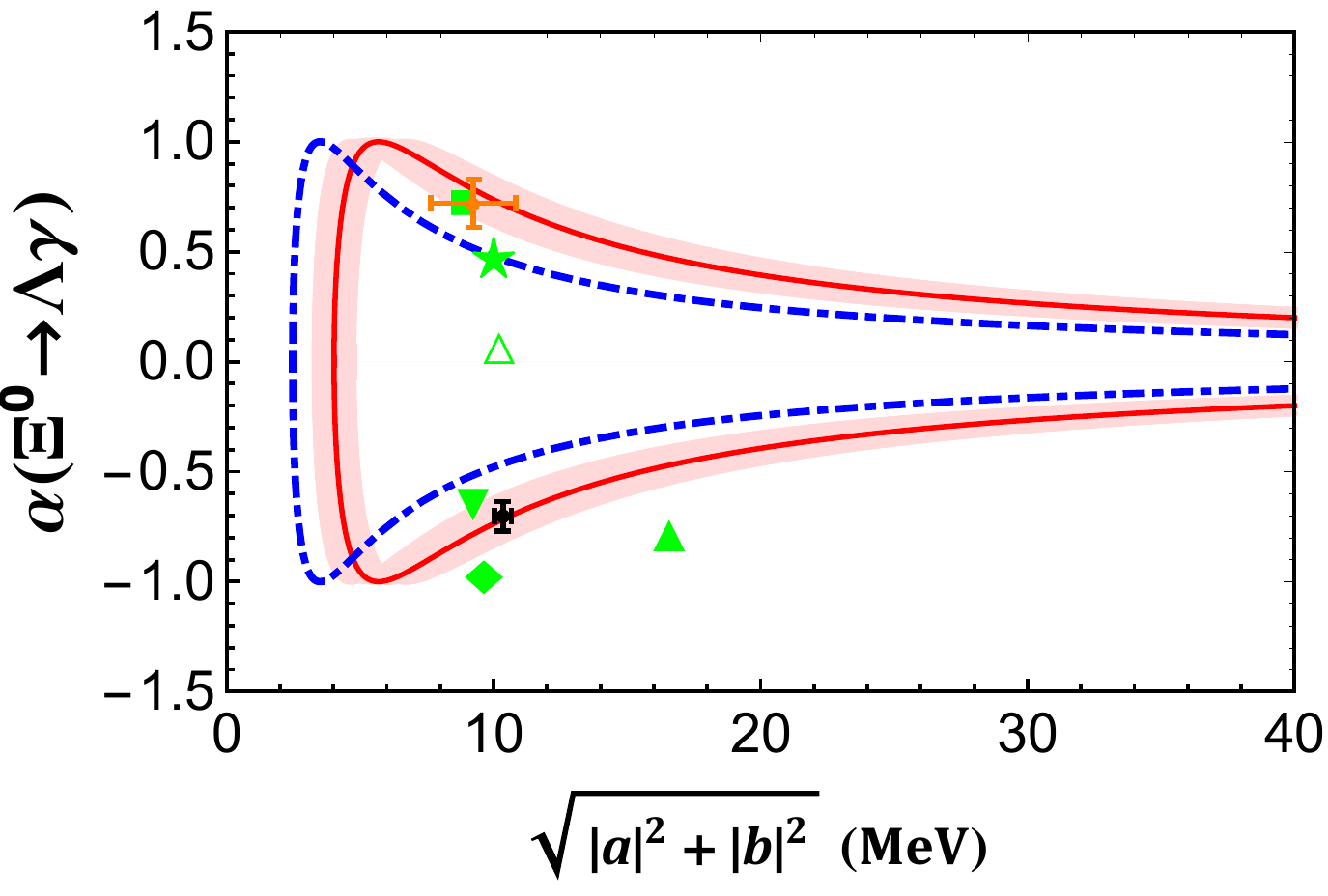}~\includegraphics[width=0.45\linewidth]{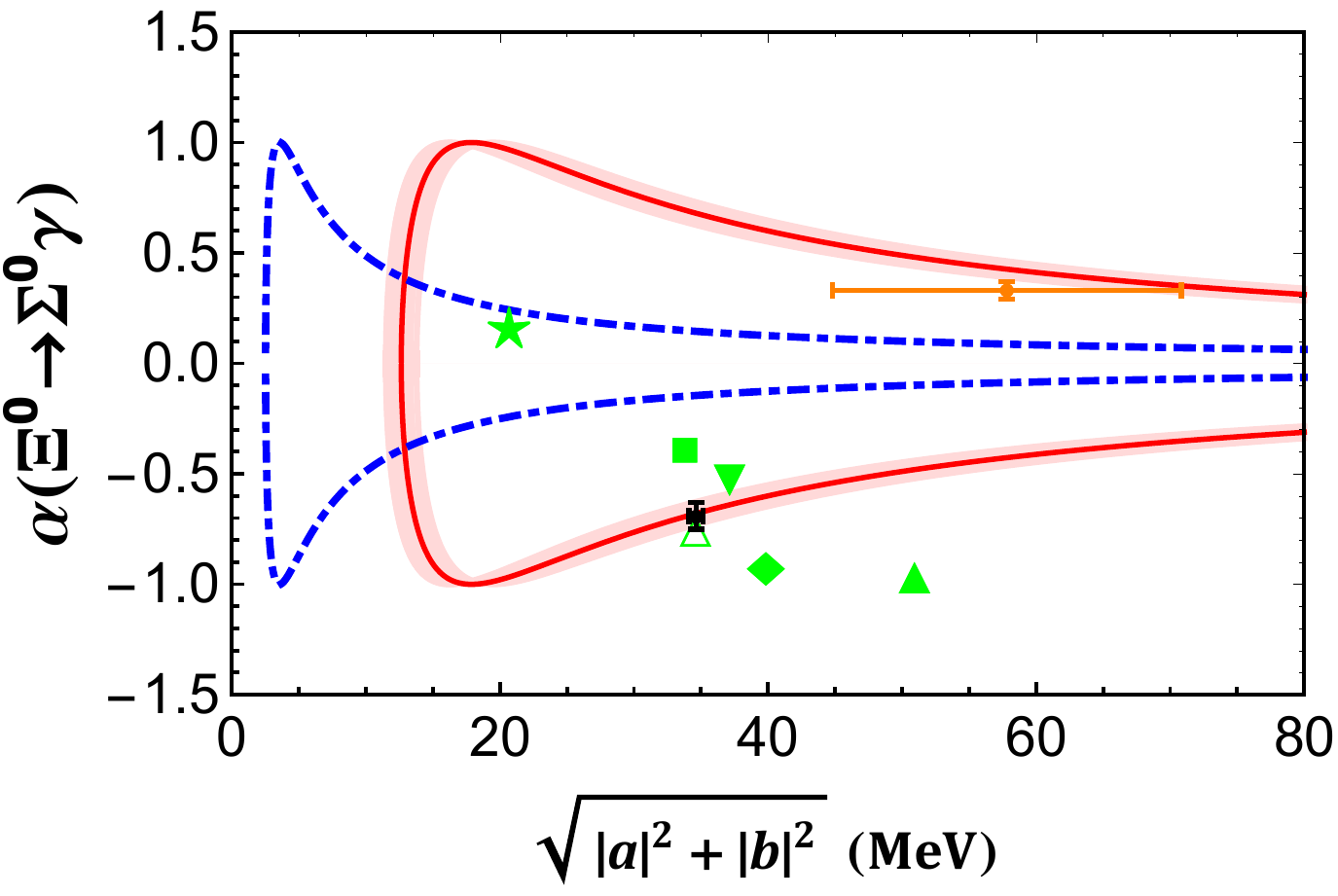}\\
  
  \caption{Same as Fig.~\ref{Fig:PreObsLambton} but for the $\Xi^0\to\Lambda^0\gamma$ and $\Xi^0\to\Sigma^0\gamma$  decays.}\label{Fig:PreObsXi0}
\end{figure*}
In Fig.~\ref{Fig:PreObsXi0}, we plot the asymmetry parameters of the $\Xi^0\to\Lambda\gamma$ and $\Xi^0\to\Sigma^0\gamma$ as functions of $\sqrt{a^2+b^2}$ in comparison with  those obtained in other approaches. It should be stressed there the loop functions donot contribute to the imaginary parts of amplitudes $a$ and $b$ of these two decays. As explained in the main text, we have fitted the real parts of amplitudes $a$ and the LEC $C_\rho$ to the experimental branching fractions and asymmetry parameters. As a result, the red solid curves are obtained with the so-determined $C_\rho$ but allowing for the real parts of amplitudes $a$ to vary.

\end{document}